\documentclass[fleqn,usenatbib]{mnras}

\usepackage{newtxtext,newtxmath}

\usepackage[T1]{fontenc}

\DeclareRobustCommand{\VAN}[3]{#2}
\let\VANthebibliography\thebibliography
\def\thebibliography{\DeclareRobustCommand{\VAN}[3]{##3}\VANthebibliography}

\usepackage{graphicx}	
\usepackage{amsmath}	
\usepackage{gensymb}
\usepackage{multirow}
\usepackage{subcaption}
\usepackage{ulem}

\title[Interstellar Near Earth Objects]{Close Encounters of the Interstellar Kind: Exploring the Capture of Interstellar Objects in Near Earth Orbit}

\author[Mukherjee et al.]{
Diptajyoti Mukherjee$^{1}$\thanks{E-mail: diptajym@andrew.cmu.edu (DM)},
Amir Siraj$^{2}$,
Hy Trac$^{1,3}$,
and Abraham Loeb$^{2}$
\\
$^{1}$McWilliams Center for Cosmology, Department of Physics, Carnegie Mellon University, 5000 Forbes Avenue, Pittsburgh, PA 15213, USA\\
$^{2}$Department of Astronomy, Harvard University, 60 Garden Street, Cambridge, MA 02138, USA\\
$^{3}$NSF AI Planning Institute for Physics of the Future, Carnegie Mellon University, 5000 Forbes Avenue, Pittsburgh, PA 15213, USA
}

\date{Accepted XXX. Received YYY; in original form ZZZ}

\pubyear{2015}

\begin{document}
\label{firstpage}
\pagerange{\pageref{firstpage}--\pageref{lastpage}}
\maketitle

\begin{abstract}
Recent observations and detections of interstellar objects (ISOs) passing through the solar system have sparked a wave of interest into these objects. Although rare, these ISOs can be captured into bound orbits around the Sun. In this study, we investigate the novel idea of capture of ISOs into near-Earth orbits and find that a steady population of ISOs exists among the current population of Near Earth Objects (NEOs).
Using numerical simulations, we find that the capture of ISOs into near-Earth orbits is dominated by Jupiter which is $10^4\times$ more efficient in capturing ISOs compared to Earth. Captured ISOs are more likely to be in orbits with high eccentricities and low inclinations. We also investigate the stability of captured ISOs and find that they are generally unstable and have an average survival life time of $\sim 1$ Myr, consistent with lifetime of NEOs originating from outer asteroid belt, and are ejected from the solar system due to interactions with other planets or the Sun.
Our results have important implications for understanding the population of ISOs in the solar system and possible future detection. We find that about one to a few $50-70$ m sized captured ISOs among NEOs would be detectable by LSST over its lifetime. By detecting and studying captured interstellar objects, we can learn about the properties and origins of such objects, and the formation and evolution of exoplanetary systems and even our solar system.

\end{abstract}

\begin{keywords}
celestial mechanics -- asteroids: general -- minor planets, asteroids: individual: 1I/‘Oumuamua 
\end{keywords}

\graphicspath{{./}{figures/}}



\section{Introduction} \label{sec:intro}
The discovery and identification of interstellar objects (ISOs) 1I/'Oumuamua \citep[e.g.,][]{Meech2017Natur.552..378M, Micheli2018Natur.559..223M}  and 2I/Borisov \citep[e.g.,][]{Jewitt2019ApJ...886L..29J} recently has led to a flurry of interest in the origins and dynamics of these objects. \cite{Siraj2022ApJ...939...53S, Siraj2022ApJ...941L..28S} also reported the discovery of two apparent interstellar meteoroids CNEOS 2014-01-08 (IM1) and CNEOS 2017-03-09 (IM2) which produced fireballs that were detected by satellites. 
Alternately, recent theories have shown that the data for IM1 to be in agreement with a stony meteroid originating in the Solar System \citep[e.g.,][]{Brown2023arXiv230614267B}. Several  hypotheses have been put forward to explain the dynamical origins of such objects. For example, \cite{Simon2018MNRAS.479L..17P, Simon2021A&A...647A.136P} and \cite{Torres2019A&A...629A.139T} considered the possibility of interstellar interlopers as objects that had been ejected from the debris disk due to encounters with other stars in the birth cluster of the star. \cite{MoroMartin2018ApJ...866..131M,MoroMartin2019AJ....157...86M} also considered the possibility that such objects originated in  young protoplanetary disks or were a part of an exocometary cloud. \cite{Hands2019MNRAS.490...21H} explored the exchange of these objects between different stellar systems in an open cluster environment. \cite{Cuk2018ApJ...852L..15C} found that the origins of such objects were better explained as a tidal disruption fragment from a binary star system. It is, therefore, imperative to study ISOs as they can open up avenues for understanding origin and dynamics of objects beyond the solar system \citep[e.g.,][]{Seligman2020ApJ...896L...8S,Hoang2020ApJ...899L..23H,Hoang2023ApJ...951L..34H,Siraj2022AsBio..22.1459S,Bergner2023Natur.615..610B}. 

Although ISOs are rare, their discovery suggests that they could be captured by planetary systems. The capture of interstellar objects by the solar system has been studied for a long time \citep[e.g.,][]{Valtonen1982ApJ...255..307V,Torbett1986AJ.....92..171T}. Most studies have explored the cross-section of capture of the Sun-Jupiter system and captures as a result of close encounters with the giant planets \citep[e.g.,][]{Siraj2019ApJ...872L..10S,Hands2020MNRAS.493L..59H,Napier2021PSJ.....2...53N,Dehnen2022MNRAS.512.4062D,Dehnen2022MNRAS.512.4078D}. Jupiter is expected to dominate the capture cross section and the rate of capture in the present day solar system is expected to be low due assuming that the hyperbolic excess velocity of the ISOs is similar to the field star velocity distribution \cite[e.g.,][]{Napier2021PSJ.....2...53N,Dehnen2022MNRAS.512.4078D}. However, the capture of ISOs would be easier when the Sun was still residing in its birth cluster given the low velocity dispersion of field stars in the birth cluster.

The question of whether any captured ISOs reside in the solar system currently is of major interest and is a matter of investigation. \cite{Morbidelli2020MNRAS.497L..46M} report that no known planetesimals of interstellar origin currently reside in the solar system. \cite{Napier2021PSJ.....2..217N} simulated a suite of 270k synthetic captured objects and found that most captured objects were not able to raise their perihelion efficiently leading to repeated scatterings with the giant planets which make them unbound over time. Only 3 out of 270k objects simulated survived for a duration of 1 Gyr. Therefore, to constrain any useful information about any possible population of interstellar objects in the current solar system, more simulations are required. 

Due to the larger number density of smaller interstellar objects, we would except smaller captured ISOs to outnumber any larger ISOs, should they be present. The smaller objects would be harder to detect unless they were close to Earth. This compels us to ask an interesting question: does a population of ISOs lie amongst the Near Earth Objects (NEOs) today?  Although the capture rate of ISOs is expected to be dominated by the giant planets, especially Jupiter, Earth can play a role as well in capturing ISOs into NEO orbits. An interesting but unexplored avenue of exploration is related the capture of ISOs by the Earth-Moon system. We expect the number of smaller-sized ISOs to be much larger than the number of 'Oumuamua and Borisov sized ISOs \citep[e.g.,][]{Siraj2022ApJ...939...53S}. This would ensure that despite the smaller capture cross-section, the Earth-Moon system would be able to capture some of these smaller sized objects. The orbital parameters of such a population would need to be identified and compared to that of present day NEOs. If a population of such captured objects exist among present day NEOs, they could be more easily detectable and can be visited by space probes which would give us invaluable information about the formation and evolution of exoplanetary systems and even our own solar system. This compels us to study whether interstellar NEOs exist and therefore motivates this work.

In this study, we explore the dynamics and capture of objects by the Earth-Moon and Jupiter systematically using a large suite of $N$-body scattering simulations. The total number of scattering events simulated is $\sim 10^{11}$ which can be considered state-of-the-art. This is required given the small capture cross-section of the Earth-Moon system. We also explore the role of the Moon on capture of ISOs by the Earth. In addition, we compare the dynamics of captured objects by the Earth-Moon system with that of a giant planet like Jupiter and explore whether close encounters with Jupiter can lead to production of any NEOs. After performing long-term simulations, we discuss the possibility of finding ISOs among a population of NEOs today.

We begin by introducing a new hybrid integration scheme that enables us to perform a large number of simulations in Section \ref{sec:methods}. This is followed by a discussion of  our models and initial conditions in Section \ref{sec:models}, and results in Section \ref{sec:results}. We, then, discuss our results in the context of previous works and compare our captured population to the population of small bodies in the Solar System at present day in Section \ref{sec:discussion}. We conclude in Section \ref{sec:conclusions} and talk about some future work that needs to be done to further this research.

\section{Methods} \label{sec:methods}

In order to derive robust statistical results, a large number of scattering experiments need to be performed. A full numerical solution involves numerically integrating the orbit of the planet (Earth/Jupiter) and any associated Moons taken into consideration and the orbit of the incoming ISO. Since the trajectory of the incoming ISO is far from the orbit of the planet at most times, a full numerical solution becomes quite inefficient since the orbits of the planets are approximately Keplerian. This presents a computational bottleneck. In order to alleviate this issue, we have derived a new hybrid method that is much faster than full numerical solution while producing similar results as that of the latter. 

Since the scatterings occur over a time-scale that is much shorter than the secular timescale of the system and the incoming test particle has zero mass, we can model the orbit of Earth-Moon/Jupiter around the Sun analytically as a pure Keplerian orbit. In order to do it, we split the system hierarchically into two subsystems. For Earth-Moon, the movement of the Earth-Moon barycenter around the sun is one subsystem whereas the movement of the Moon around the Earth is another subsystem. For Jupiter, we don't consider any associated moons and thus we only evolve its orbit. For the timescales in question for the scattering experiments, we compare the motion of the system using our hierarchically split Keplerian method to those from full numerical results and find similar results. The positions and velocities of the particles in the system are generally agree within $10^{-8}-10^{-10}$ of each other. We use a Kepler solver that uses the universal variable formulation \citep[][]{Wisdom2015MNRAS.453.3015W}.

To integrate the motion of the test particle under the influence of the Sun-Earth-Moon system (or any other system for that matter), we use a time-symmetrized fourth-order Hermite scheme\citep[][]{Kokubo1998MNRAS.297.1067K}. 

We perform long-term integrations for a subset of captured objects. For these simulations, we include every planet except Mercury to understand the lifetime of captured ISOs in NEO orbits. We again use the Kepler solver to integrate the orbits of the planets without taking into account any interactions between the planets. The Keplerian portion of the integration is done in Jacobi coordinates \citep[e.g.,][]{Murray2000ssd..book.....M,Rein2015MNRAS.452..376R}. Essentially our  integrator then reduces to a Wisdom-Holman \citep[][]{Wisdom1991AJ....102.1528W} like scheme without the interaction terms between planets taken into account. Although we expect the interaction terms to contribute to the secular evolution of the planets, we expect the effect of the secular evolution of the planetary orbits on the orbit of the ISO to be minimal. We stress that the usage of the Hermite integrator along with the Keplerian scheme allows us to simulate close encounters which are necessary for the long term simulations.

We time-symmetrize the integrator by using an iterative $P(EC)^3$ scheme \citep{Kokubo1998MNRAS.297.1067K} along with symmetrized timesteps as used in \cite{Pelupessy2012NewA...17..711P}. We use a globally adaptive timestep where the timesteps are determined only by the scattered particle. Time steps are taken as the minimum of the following timesteps:

\begin{gather}
\tau_{\rm freefall} = \eta \sqrt{\frac{r_{ij}} {a_{ij}}} \\
\tau_{\rm flyby} = \eta \frac{r_{ij}} {v_{ij}}
\label{eq:timestep_criterion}
\end{gather}
where $r_{ij}$ is the relative distance, $v_{ij}$ is the relative velocity, and $a_{ij}$ is the relative acceleration between two particles. $\eta$ is tunable parameter that controls the accuracy of integration. We set it to $10^{-3}$ in our simulations.

The Hermite-Kepler hybrid integration scheme proves to be extremely efficient at simulating the scattering of ISOs and their long term evolution. We compare our results to those obtained using the IAS15 integrator \citep{Rein2015MNRAS.446.1424R} for a set of $10^7$ scattering events. We notice that the our method provides statistically similar results as that from IAS15 (see appendix). The experiments are $\sim 10-50\times$ faster when using this method compared to {\tt\string IAS15} enabling a larger number of scattering experiments to be performed. Our hybrid integrator also offers the ability to handle close encounters unlike a traditional Wisdom Holman scheme, allowing us to examine the long term evolution of captured ISOs.

\section{Modeling the scattering of objects} \label{sec:models}
\begin{figure*}
\centering
\begin{subfigure}[b]{0.45\textwidth}
\includegraphics[width=1.0\textwidth]{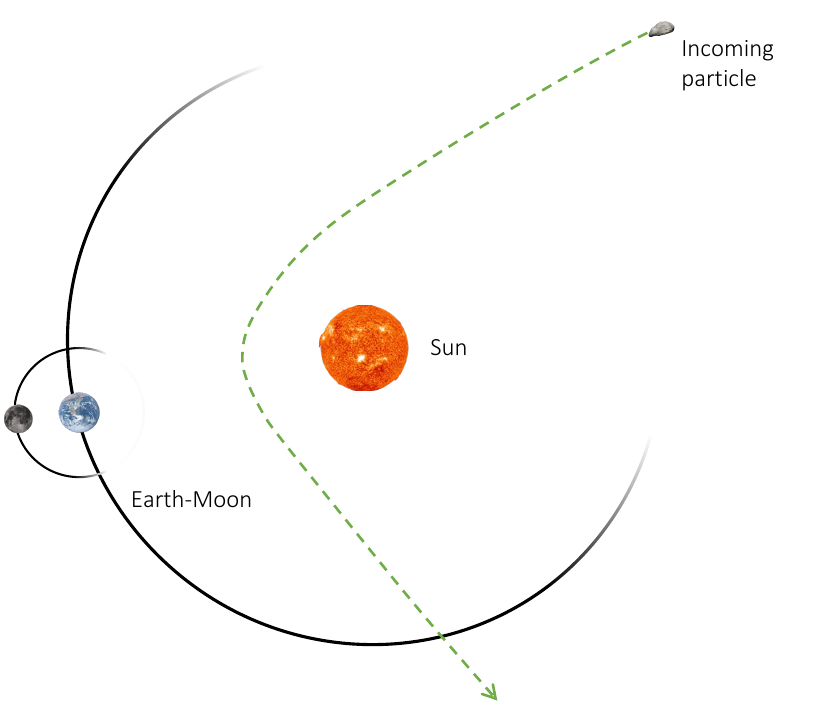}
\end{subfigure}
\hfill
\begin{subfigure}[b]{0.45\textwidth}
\includegraphics[width=1.0\textwidth]{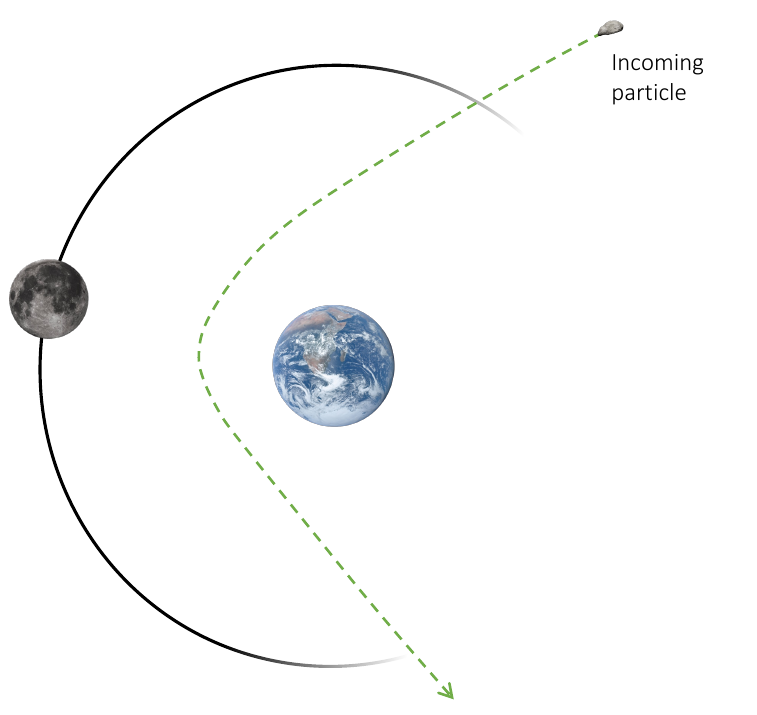}
\end{subfigure}

\caption{Visualization of trajectories of incoming particles getting scattered by the Sun-Earth-Moon system. Since there are two binary systems present, we can calculate the cross section of the whole system and that of the Earth-Moon system. \textit{Left}: A particle experiencing a close encounter with the Sun. \textit{Right}: A particle experiencing a close encounter with the Earth-Moon system. Image credit: NASA.}

\label{fig:visual}
\end{figure*}

To model an incoming interstellar object on an initially unbound orbit, we follow the steps outlined in \cite{Napier2021PSJ.....2...53N} and \cite{Dehnen2022MNRAS.512.4062D} with some differences. We consider scattering events with Earth-Moon and Jupiter. A visualization of the scattering process with the Earth-Moon is provided in Figure \ref{fig:visual}. The process of generating initial conditions is similar in both cases and we present an overview of our methods using the example of Earth-Moon before moving on to Jupiter.

In the first scenario, our system only consists of the Sun, and the Earth-Moon system. This is done in order to study the capture cross section and the influence of only the Earth-Moon system without that of the giant planets. We first explore the cross-section of the entire Sun-Earth system before moving on to simulate trajectories that undergo close encounters with the Earth-Moon binary. The initial conditions for the incoming particles are generated in different ways in the two scenarios as is detailed below.

Because of the time dependent potential of the Earth-Moon system, ISOs that pass within the orbit of the Earth around the Sun can be scattered and captured. This determines the cross-section of capture for the Sun-Earth system. In order to calculate this cross-section, we follow steps as outlined in \cite{Napier2021PSJ.....2...53N} to generate the initial conditions. 

As in \cite{Napier2021PSJ.....2...53N}, the incoming particle is sampled from a sphere with a radius of $10^9$ au from the barycenter of the Sun-Earth-Moon system. Since we are interested exploring the cross-section as a function of the hyperbolic excess velocity $v_{\infty}$, we simulate multiple scattering events for a particular $v_{\infty}$. Once $v_{\infty}$ is chosen, the speed of the particle at the current radius ($v_{\mathrm{r}}$) can be derived using the conservation of energy as 

\begin{equation} 
    v_{\mathrm{r}} = \sqrt{v_{\infty}^2 +   \frac{2G\mu}{d_{\mathrm{bary}}}}
    \label{eq:cons_energy}
\end{equation}
where $\mu$ is the barycenteric mass of the Sun-Earth-Moon system and $d_{\rm {bary}}$ is the distance of the particle from the barycenter of the system. The velocity vector for the particle is then found by multiplying the velocity from equation \ref{eq:cons_energy} with the direction directly towards the barycenter of the system.

Given a choice of $v_{\infty}$, the maximum impact parameter ($b_{\mathrm{max}}$) of the object can be derived. The maximum impact parameter is given as 
\begin{equation}
    b_{\mathrm{max}} = \sqrt{q_{\mathrm{max}}^2 + 2q_{\mathrm{max}} |a|}
    \label{eq:bmax}
\end{equation}
where $q_{\mathrm{max}}$ is the maximum perihelion distance sampled and $a$ is the semi-major axis of the incoming particle. We set $q_{\mathrm{max}}=1.3$ au in our simulations. This is mainly motivated by the definition of NEOs in literature where objects with perihelion $\leq 1.3$ au are considered to be NEOs \citep[e.g.,][and references therein]{Morbidelli2002aste.book..409M}.
The impact parameter of the object is sampled uniformly between $0$ and $b_{\mathrm{max}}$ and the object is placed on tangent plane to the sphere at the point in a random direction. Some previous studies have sampled the impact parameter assuming a probability distribution $\propto b^2$ \citep{Siraj2019ApJ...872L..10S,Hands2020MNRAS.493L..59H,Dehnen2022MNRAS.512.4062D}. In such a scenario, more encounters take place closer to the planet rather than far from it. We ran a set of simulations assuming the distribution of impact parameters was $\propto b^2$ and found no major differences in the computed cross-section. 

 To save on computation time, we then evolve the motion of the particle and the barycenter of the system as a pure Keplerian orbit until the particle has reached a distance of $20$ au from the barycenter. This value is similar to the initial starting distance used in \cite{Dehnen2022MNRAS.512.4062D} but lower than that in \cite{Napier2021PSJ.....2...53N}. We verified that the lower value used in our simulations do not affect the results. 

Once the particle has reached a distance of $20$ au, we evolve the particle under the influence of the Sun and the Earth-Moon system. We put in the Earth-Moon system around the Sun. The positions of the Earth and Moon are generated with random orbital phases with the orbital phase being drawn uniformly between $0-2\pi$. This is done to prevent any spurious results that may arise out of any specific configuration of orbital phases. The system with the test particle is then evolved until the particle has been scattered off by the system. In case the specific energy of the particle drops below zero, we stop the simulation and store the positions and velocities of all particles in the simulation. In case of collision with any of the other objects in the system, we also record the positions and velocities of all particles in the system. If the outgoing test particle is unbound after it reaches a distance $> 20$ au after scattering, we stop the simulation and conclude that the particle could not be captured. For all of the captured particles, the orbit is evolved until the particle has reached a distance $> 20$ au and then the orbital parameters are calculated.

To generate trajectories that undergo close encounters with the Earth-Moon, we follow the steps above and then select orbits that undergo close encounters with the Earth-Moon system, similar to \cite{Dehnen2022MNRAS.512.4062D} with some differences. We define a close encounter when the minimum distance between the incoming ISO and the Earth is lower than than the Hill radius of Earth (0.01 au),similar to within L1 and L2.

To select the subset of systems wherein the incoming ISO undergoes close encounter with the Earth-Moon, we chose only those orbits where the hyperbola of the orbit of the incoming ISO intersects with the orbit of the Earth-Moon system and lies within the Hill radius of Earth. 
This is a non-trivial problem as it involves finding the Minimum Orbit Intersection Distance (MOID) and selecting a subset where the MOID is less than $0.01$ au. To accomplish this, we use a fast, geometric method to find the MOID using the technique of \cite{Moid2013AcA....63..293W}. The calculation is accurate to $10^{-14}$ au and any missed MOIDs do not affect our results. 

Once an orbit has been found, we evolve the particle on this hyperbola, and place the Earth-Moon system at a random orbital phase such that the incoming ISO is at the MOID distance with respect to Earth at $t=2$ yr. The orbit of the ISO is simulated for a total for $t=5$ yr after which the specific energy of the ISO is evaluated to check whether it was captured or not. Collisions with the Earth-Moon system are also tracked. An additional set of simulations without the Moon is performed using the same initial conditions. This is done in order to study the effect of the Moon on captures by close encounters with the Earth-Moon system.

We also perform a set of $10^9$ simulations with the Sun-Jupiter system in order to understand how the orbital properties of the captured object differs when captured by the Earth-Moon system versus that by Jupiter and to study whether Jupiter can efficiently capture interstellar NEOs.  The process of generating initial conditions for ISOs scattered by Jupiter is similar to that mentioned above. For the Sun-Jupiter system, we perform a full integration once the particle has reached a distance $<100$ au. $q_{\mathrm{max}}$ is set to 7.3 au for these Sun-Jupiter simulations.

To study the lifetime of the captured objects, we perform long-term simulations upto $10$ Myr using the hybrid scheme described in the previous section. Once the scattering simulations have ended, we place the captured object and the planetary system at their respective positions in the solar system. Our solar system consists of all planets except Mercury whose effect is expected to be minimal. Since the long-term simulations are computationally expensive even with the hybrid integration scheme, we restrict ourselves to performing integrations of $\approx 20$k captured objects with $a<20$ au that are randomly selected.

\section{Results} \label{sec:results}
\subsection{Capture cross section of the Sun-Earth-Moon system} \label{subsec:sunearth_capture}

We performed a total of $5\times10^{10}$ simulations to calculate the cross-section of capture of the Sun-Earth system. More simulations were performed for simulations with higher $v_{\infty}$ because of the smaller capture cross section. To calculate the cross-section of capture, we used the following formula \citep{Dehnen2022MNRAS.512.4062D}:

\begin{equation}
    \sigma = \pi b_{\mathrm{max}}^2 \frac{N_{\mathrm{captured}} \pm \sqrt{N_{\mathrm{captured}}}}{{N_{\mathrm{sampled}}}}
    \label{eq:sigma_calc}
\end{equation}

To calculate $N_{\mathrm{captured}}$, we only consider particles with semi-major axes $a < 10^5$ au. Particles with larger values of $a$ can be stripped by nearby stars and become unbound from the system. We also investigated using a different formula for calculating $\sigma$ using equation (29) from \cite{Napier2021PSJ.....2...53N} but found minimal differences.

\begin{figure*}
    \begin{center}
    \includegraphics[width=1.0\textwidth]{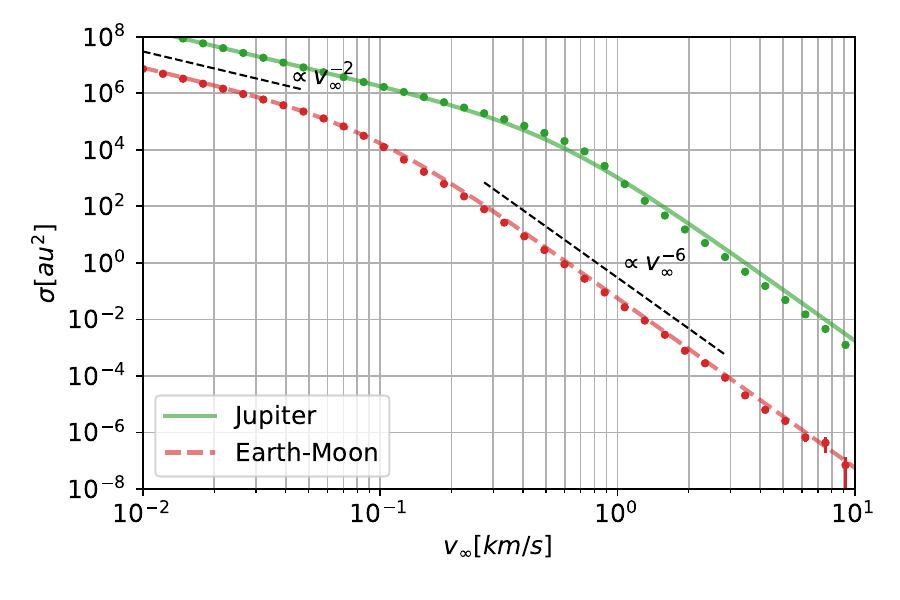}
    \caption{Capture cross section ($\sigma$) presented as a function of the hyperbolic excess velocty ($v_{\infty}$). We have compared the cross section of the Earth-Moon to that of Jupiter. The dots represent the actual data points calculated from the simulation and the lines represent the fitted cross section. We note that the cross section of Earth-Moon follows the same qualitative form as that of Jupiter. However, the cross-section of the Earth-Moon system is 1\% of that of Jupiter at lower velocities whereas it is 0.01\% that of Jupiter at higher velocities.} 
    \label{fig:cross_section_full}
     \end{center}
\end{figure*}

Looking at Figure \ref{fig:cross_section_full} where we plot the capture cross section ($\sigma$) as a function of $v_{\infty}$ for both Earth-Moon and Jupiter, we find that $\sigma$ behaves as $v_{\infty}^{-2}$ at lower values of $v_{\infty}$ and as $v_{\infty}^{-6}$ at higher values. This is qualitatively similar to the results from \cite{Napier2021PSJ.....2...53N}. We fit a function of the form 

\begin{equation} \label{eq:fitting_equation}
    \sigma_{\mathrm{fit}} = \frac{\sigma_0}{ \left (\frac{v_{\infty}}{v_{0}} \right)^2 \left[ \left (\frac{v_{\infty}}{v_{0}} \right)^2 + 1 \right]^2}
\end{equation}
to the measured datapoints and find that $\sigma_0 = 9.825 \times 10^4 \pm 3.55 \times 10^3 \mathrm{au}^2$ and $v_{0} = 0.0914 \pm 0.0011 \mathrm{km/s}$. The latter of these values represents the velocity at which the capture cross section transitions from a $v_{\infty}^{-2}$ form to a $v_{\infty}^{-6}$ form. We calculate the cross-section of Jupiter in a similar fashion by performing $10^9$ scattering simulations and compare it to that of Earth-Moon system. Although they follow the same functional form, Jupiter transitions from a $v_{\infty}^{-2}$ form to a $v_{\infty}^{-6}$ form at higher $v_{\infty}$. Indeed we find that $v_{0,\mathrm{jup}} = 0.554 \pm 0.020 \mathrm{km/s} $. We find that $\sigma_{0,\mathrm{jup}}=6.09\times10^4 \pm 4.47 \times 10^3 \mathrm{au^2}$. The cross-section of capture for the Sun-Earth system is 1\% of that of Jupiter at smaller $v_{\infty}$ and $0.01$\% that of Jupiter at larger $v_{\infty}$. Although our fitted values are slightly different from that found by \cite{Napier2021PSJ.....2...53N}, we note that our fitted cross-section only differs by $\leq 10\%$ compared to theirs. This difference can be attributed to the presence of other giant planets like Saturn while calculating the capture cross section in \cite{Napier2021PSJ.....2...53N}. However, the similarity in the results pertaining to the capture cross section of Jupiter validates the robustness of our computational methods.

\begin{figure*}
\centering
\begin{subfigure}[b]{\linewidth}
\includegraphics[width=\textwidth]{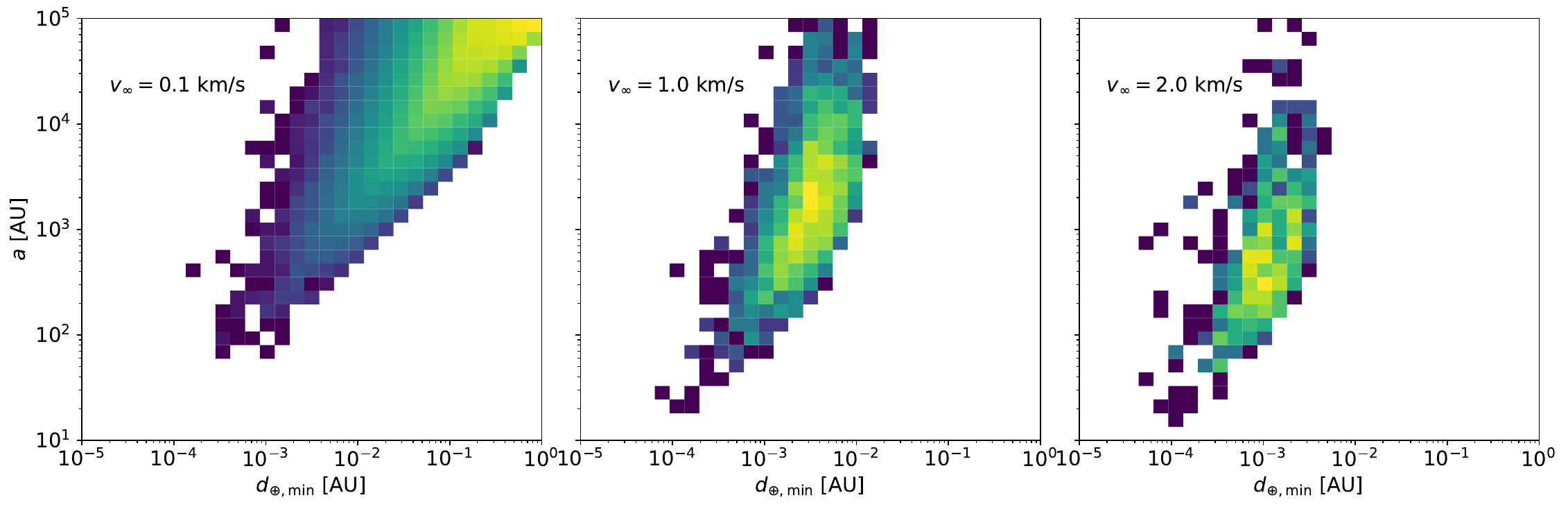}
\end{subfigure}

\begin{subfigure}[b]{\linewidth}
\includegraphics[width=\textwidth]{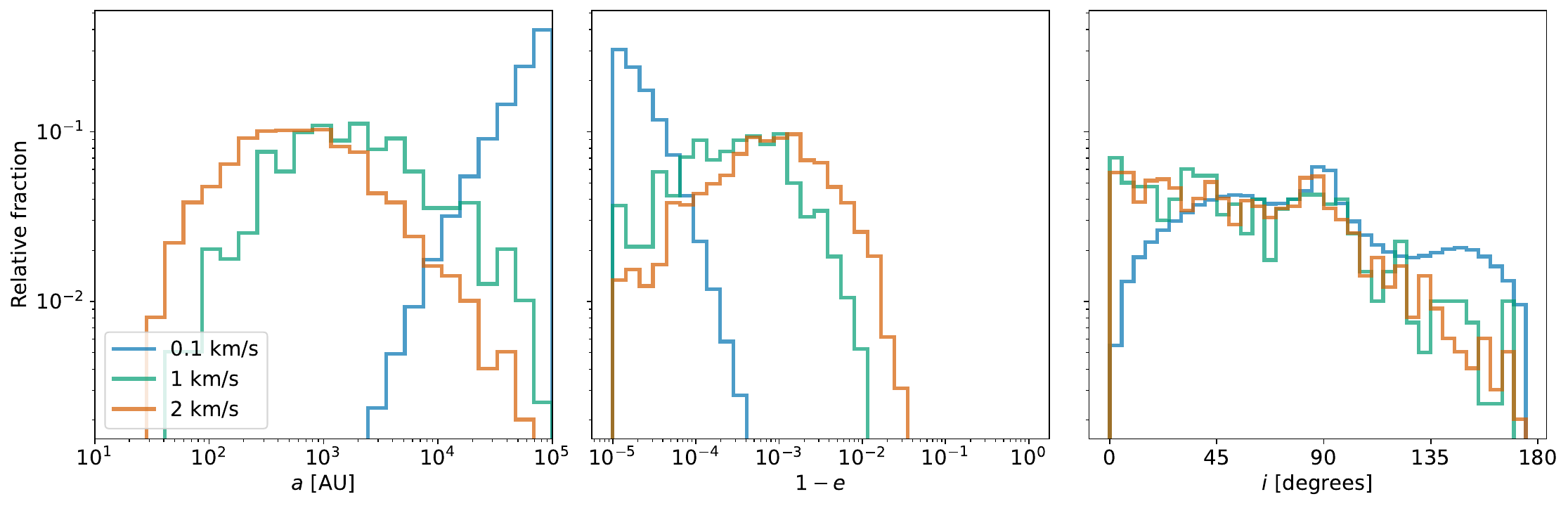}
\end{subfigure}

\caption{Distribution of the orbital parameters of the objects captured by the Sun-Earth system. \textit{Top}: Distribution of the minimum distance to Earth $d_{\oplus,\mathrm{min}}$ and the captured semi-major axis $a$. The brighter colors denote larger number of captued objects whereas darker colors indicate fewer. At $v_{\infty} = 0.1$ km/s, the dominant source of captures are distant encounters. As $v_{\infty}$ increases, more captures occur because of close encounters with the Earth-Moon system. In fact, all captures for $v_{\infty} \geq 1$km/s occur due to close encounter with the Earth-Moon system. \textit{Bottom}: Semi-major axis ($a$), eccentricity($e$), and inclination ($i$)  distribution of captured objects. As $v_{\infty}$ increases, the median of the semi-major axis distribution decreases. We also find that prograde orbits are preferred for captures.    
\label{fig:orbital_params_full}}
\end{figure*}

Objects that are captured can have varying pericenter distances with respect to the planet. We define distant encounters as those where the minimum distance of the incoming object to the planet is greater than the Hill radius of the planet and close encounters as encounters where the minimum distance is within the Hill radius of the planet. Both distant and close encounters can result in the capture of ISOs. To understand which type dominates as a function of the hyperbolic excess velocity, we plot the distribution of particles as a function of the minimum distance to Earth $d_{\oplus,\mathrm{min}}$ and the semi-major axis $a$ of the captured objects in Figure \ref{fig:orbital_params_full} We find that at lower $v_{\infty}$ distant encounters dominate the capture. Distant encounters, in fact, account for $\approx 99.7\%$ of the captures  at low hyperbolic excess velocities. The ISOs captured due to distant encounters tend to have larger semi-major axes and smaller inclinations. The median semi-major axis of objects captured due to distant encounters at $v_{\infty}=0.1$ km/s is $6\times10^4$ au, whereas that for close encounters is $2.2\times10^3$ au. The median inclination for distant encounters is $59 \degree$ while that for close encounters is $98 \degree$. At $v_{\infty} \geq 1$ km/s, close encounters become the dominant mechanism to capture ISOs. Close encounters permit captures of objects with larger inclinations and at smaller $v_{\infty}$ even objects in retrograde orbits can be captured.

Next, we analyze the orbital parameters of the captured objects in Figure \ref{fig:orbital_params_full}. 
As $v_{\infty}$ increases, looking at Figure \ref{fig:orbital_params_full}, we find that the distribution of the semi-major axis of captured objects moves towards lower values. The median of the distribution at $v_{\infty}=0.1 $km/s is $6\times 10^4$ AU whereas that for $v_{\infty}=2.0 $km/s is about $5.5\times 10^2$ AU. This leads to a fatter tail in the eccentricity distribution as well. At higher $v_{\infty}$ most captures take place on orbits that are prograde with respect to the orbit of the Earth around the Sun. Orbits on lower inclinations allow for longer interaction time between the incoming ISO and the Earth-Moon system leading to easier captures. There is a pronounced peak at $i=90 \degree$ but this is caused due to the initial distribution of inclinations in our system. 

\subsection{Close Encounters with Earth-Moon System}
The fraction of orbits that undergo close encounters with the Earth-Moon is minuscule in the simulations presented in \ref{subsec:sunearth_capture}. As we saw in the previous section, close encounters are the dominant source in production of captured objects at larger $v_{\infty}$ and objects that have smaller $a$. Since we are interested in objects that have been captured into the solar system ($a < 30$ au), we expect close encounters with the Earth-Moon to be the dominant mechanism rather than distant encounters.  Here we present a closer examination of the captures resulting from close encounters and the cross-section of capture into smaller semi-major axis.

Since close encounters occur less frequently, the method of previous section proves to be extremely computationally expensive to obtain better statistics on objects captured due to such encounters. 
Therefore, we perform a different set of simulations where the orbit of the incoming ISO falls within the Hill radius of the Earth using the method described in section \ref{sec:models}. We perform a set of $4\times10^{10}$ simulations to calculate the cross-section of capture of the Earth-Moon system, the orbital parameters of the captured objects, and the effect of the Moon on the capture cross-section of the Earth. 

\begin{figure*}
    \begin{center}
    \includegraphics[width=0.8\textwidth]{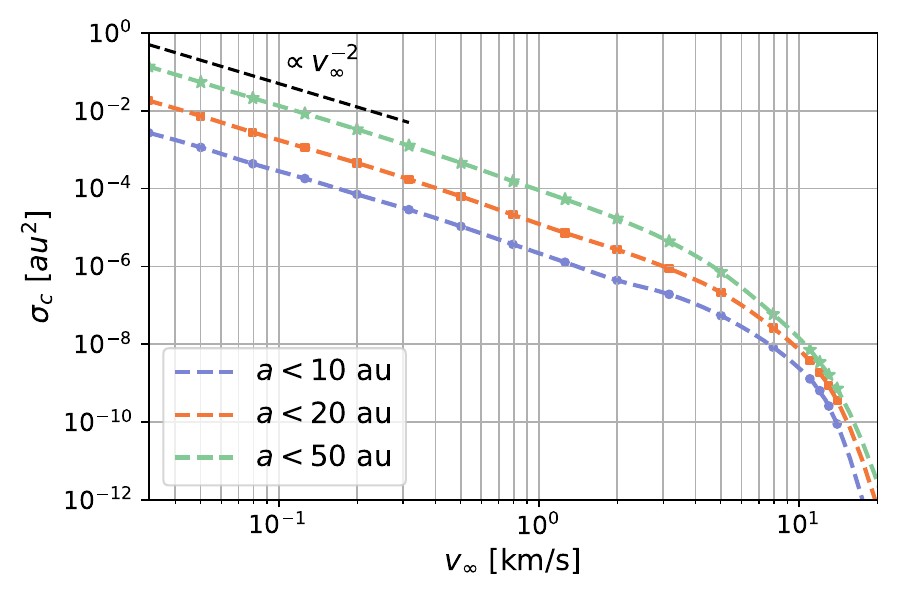}
    \caption{The capture cross section of close encounters ($\sigma_c$) with Earth-Moon system as a function of the hyperbolic excess velocity ($v_{\infty}$) for three different sets of maximum semi-major axes($a$). Since close encounters are rare, the capture cross section is much lower than the full capture cross section calculated in Figure 2. However, the capture cross-section follows the same qualitative functional profile as that in Figure 2.} 
    \label{fig:cross_section_close}
     \end{center}
\end{figure*}

Only a fraction $f_c$, where $f_c$ is given as 
\begin{equation}
    f_c = \frac{r_{\mathrm{Hill}}^2}{q_{\mathrm{max}}^2}
\end{equation}
of the total scattering events result in a close encounter. Therefore, to calculate the capture cross-section resulting from close encounters, we modify Equation (5) slightly to obtain
\begin{equation}
    \sigma_c = f_c \pi b_{\mathrm{max}}^2 \frac{N_{\mathrm{captured}} \pm \sqrt{N_{\mathrm{captured}}}}{{N_{\mathrm{sampled}}}}
    \label{eq:cross_section_close}
\end{equation}
where $N_{\mathrm{sampled}}$ is now the total number of close encounters sampled. To test the validity of this method, we compare the cross section of capture at larger $v_{\infty}$ from the previous section to the cross section obtained for objects captured with $a < 10^5$ au using this method. Since the captures at larger $v_{\infty}$ are dominated by close encounters, we should expect both methods to give us consistent values for the capture cross-section at larger $v_{\infty}$. We find that both methods produce results that are within 1\% of each other indicating this method to calculate the capture cross-section of close encounters is valid.

We present the cross section of capture due to close encounters ($\sigma_c$) calculated using equation \ref{eq:cross_section_close} in Figure \ref{fig:cross_section_close} in order to understand how the cross section changes as a result of changing the threshold semi-major axis. We calculate the cross section for three different values of the threshold semi-major axis $a_{t}$: 10 au, 20 au, and 50 au. We only select particles with semi-major axes below the threshold semi-major axis to calculate $N_{\mathrm{captured}}$. We find that the cross-section falls off as $v_{\infty}^{-2}$ until $10$ km/s and then it drops rapidly. 

Fitting a functional form to the capture cross section of close encounters $\sigma_c$ is more complicated than that presented in the previous section for all encounters. In order to do so, we assume that the cross section drops off as $v_{\infty}^{-6}$ after $v_{\infty} = 15$ km/s. Note that while the assumption might not exactly hold, it makes little practical difference since the capture cross section is dominated by smaller $v_{\infty}$ where the functional form is well known. As in the previous section, we fit a functional form similar to equation \ref{eq:fitting_equation}. However, $\sigma_0$ and $v_{0}$ are now functions of the threshold semi-major axis $a_t$ and written as $\sigma_0(a_t)$ and $v_{0}(a_t)$. We find a weak dependence of $v_{0}$ as a function of $a_t$ where $v_0(a_t)$ decreases as a function of $a_t$. However, the same is not true for $\sigma_0(a_t)$. Comparing $\sigma_0(a_t)$ as a function of $a_t$, we find that it can be best represented as 
\begin{equation}
    \sigma_0(a_t) = \sigma_{0,t} a_t^{\beta} 
\end{equation}.
We find the best fit values to be $\sigma_{0,t} \approx -1.1 \times 10^{-10} \pm 1.5\times 10^{-11} \mathrm{au}^2$ and $\beta = 3.07 \pm 0.10$. Thus, we find that $\sigma_c(a_t) \propto a_t^{3.07 \pm 0.10}$. However, we caution the reader this propotionality does not include the contribution from $v_0(a_t)$.

For $v_{\infty} > 10$ km/s, the capture cross-section drops rapidly. At $v_{\infty} = 10$ km/s, the cross-section of capture is $\sim 10^{-9} \mathrm{au}^2$. The geometric cross-section for collision with the earth is $\sigma_{\mathrm{coll}} \approx 5\times10^{-9} \mathrm{au}^2$. Thus, for  $v_{\infty}=10$ km/s or higher, collisions with the Earth dominate over captures.  

\begin{figure*}
    \begin{center}
    \includegraphics[width=1.0\textwidth]{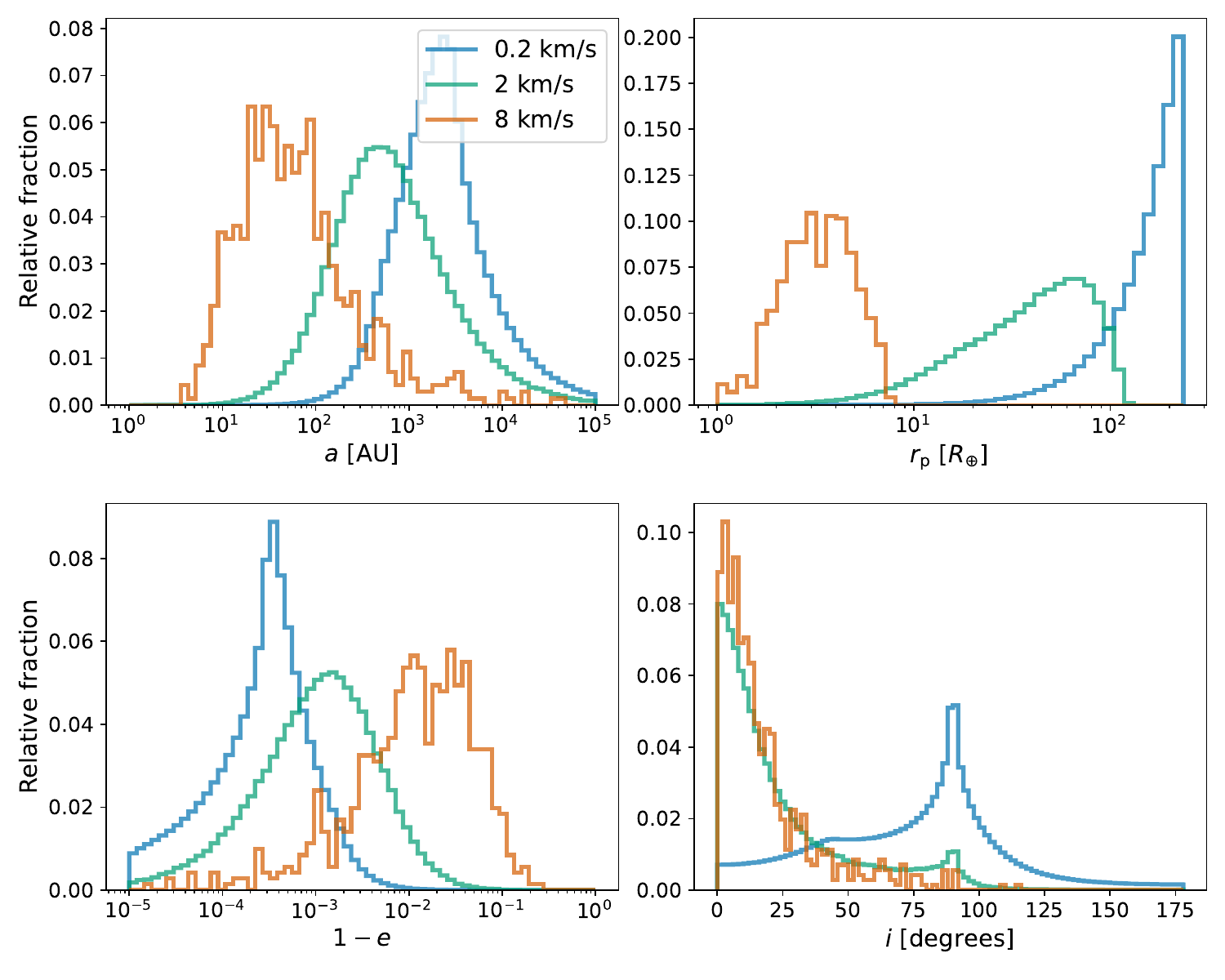}
    \caption{Orbital properties of captured objects undergoing close encounters with the Earth-Moon system and their relative fractions. \textit{Top-left}: Distribution of semi-major axis ($a$) \textit{Top-right}: Distribution of pericenter distance from Earth ($r_p$) \textit{Bottom-left}: Distribution of eccentricity ($e$) after capture. \textit{Bottom-right}: Distribution of inclination ($i$) after capture. We notice that as the hyperbolic excess velocity $v_{\infty}$ increases, the median semi-major axis $a$ and eccentricity $e$ of the captured objects decrease. Most objects are preferentially captured on prograde orbits and objects with higher velocities need closer encounters with Earth to get captured. The distributions are qualitatively similar to those in Figure \ref{fig:orbital_params_full}. We find that captures on retrograde orbits decrease significantly  for large $v_{\infty}$.} 
    \label{fig:orbital_properties_close}
     \end{center}
\end{figure*}

The distribution of orbital parameters is presented in Figure \ref{fig:orbital_properties_close} and presents an interesting story. We notice qualitatively similar trends as that from Figure \ref{fig:orbital_params_full} but we notice that the distribution of semi-major axes of captured objects skews towards smaller values than those from distant captures. As $v_{\infty}$ increases, the distribution skews towards smaller values of $a$ and in fact for $v_{\infty}=8$ km/s, the median value is 50 au. This is on account of closer approaches taking place with the Earth as is evident from the distribution of distances of closest approach to Earth ($r_{p}$). 
The distribution of inclination of the captured objects shows that with larger values of $v_{\infty}$, the inclination skews towards $i=0\degree$. There is a smaller fraction of captured objects in retrograde orbits for $v_{\infty}=0.2,2$ km/s but this drops rapidly as $v_{\infty}$ increases and we find no retrograde captures for $v_{\infty} > 8$ km/s.

An interesting examination pertains to the effect of the Moon on captures by the Earth. We run a set of simulations with the same initial orbits but with the Earth without the Moon to calculate the cross-section. We select objects with semi-major axes below a certain threshold and vary this threshold parameter. For all values of the threshold parameter $\geq 10$ au, we find little differences between the capture cross-section of the models with the Moon versus those without. However when we go to a threshold semi-major axis of $5$ au, we find certain differences at higher $v_{\infty}$. We find that the Moon actually has very little effect on the capture-section of the Earth until $v_{\infty}\sim 12$km/s. However, past that value, the capture cross section of the Earth-Moon system increases rapidly compared to the Earth-only system. At $v_{\infty}=13$ km/s, the ratio of the capture cross section with the moon  to the capture cross section without the moon is $1.3$. At $v_{\infty} > 13$ km/s models with the Moon can capture objects with $a \leq 5$ but models without the moon are unable to. At larger $v_{\infty}$, the time dependent potential of the Moon around the Earth plays a larger role in the capture of ISOs and thus leads to more captures. This effect, however, is practically negligible as the captures are dominated by captures at lower $v_{\infty}$ where the differences are negligible. 

\subsection{Captures by Jupiter and Comparison to Known Small Bodies}

\begin{figure*}
    \begin{center}
    \includegraphics[width=0.8\textwidth]{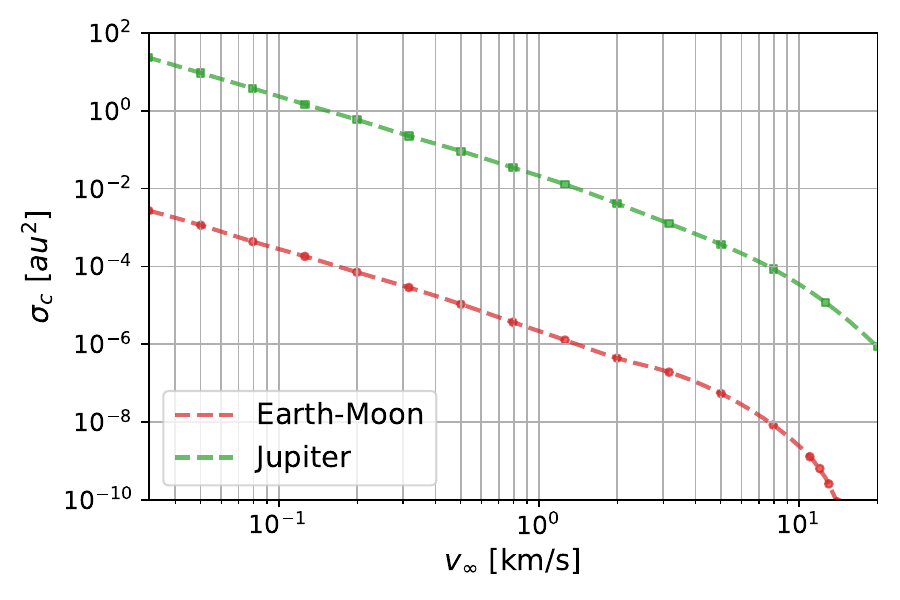}
    \caption{Capture cross section of close encounters ($\sigma_c$) of objects with $a<10$ au and $q\leq 1.3$ au as a function of the hyperbolic excess velocity ($v_{\infty}$). We note that rate of capture falls off as $v_{\infty}^{-2}$ at smaller $v_{\infty}$ and drops rapidly at larger $v_{\infty}$. Jupiter is $10^{4}$ times as efficient as Earth in injecting objects into the NEO orbits. } 
    \label{fig:cross_section_em_vs_j}
     \end{center}
\end{figure*}

Since Jupiter is expected to dominate the capture cross-section, we examine the effectiveness of Jupiter in capturing objects into near Earth orbits. For the sake of clarity, we define an object to be in a near Earth orbit if its perihelion distance $q$ is $\leq 1.3$ au \citep[e.g.,][]{Morbidelli2002aste.book..409M}. We run a set of $10^9$ simulations with the same $v_{\infty}$ as that used while simulating the Earth-Moon system and we set $q_{\mathrm{max}}=7.3$ au. We generate close encounters in a similar fashion as that mentioned before. We are primarily interested in close encounters with Jupiter. 
We calculated the efficiency of distant encounters in generating captured objects in the solar system and found that distant encounters become subdominant for large $v_{\infty}$ and $a \leq 100$ au. To capture objects on orbits with $a \leq 50 $ au, a close encounter with Jupiter is required. In this study, we consider all trajectories that pass within the Hill radius of Jupiter (0.35 au) as a close encounter.

To understand Jupiter's efficacy at injecting captured objects into NEO orbits, we consider objects with $a \leq 10$ au and $q \leq 1.3$ au and calculate the cross-section in Figure \ref{fig:cross_section_em_vs_j}. Examining the figure, we find that the capture cross-section of close encounters with Jupiter qualitatively follows the same form as that of close encounters with the Earth-Moon. However, Jupiter, owing to its mass, is $10^4 \times$ more efficient than Earth-Moon in generating NEOs from ISOs. This is not surprising since the capture cross section for planetary close encounters ($\sigma_c$) is $\propto \frac{m_p^2}{a_p}$ where $m_p$ is the mass of the planet and $a_p$ is the semi-major axis of the planet \citep[][equation 22]{Napier2021PSJ.....2...53N}. Jupiter is $\sim 300\times$ more massive than Earth and has a semi-major axis of $5.2$ au so we expect $\frac{\sigma_{c,jup}}{\sigma_{c,earth}} \approx \frac{300^2}{5.2}  \sim 2\times10^4$, which is within a factor of $2$ compared to our numerically obtained results. Performing a similar analysis as that in previous sections,we find that for Jupiter $\sigma_c(a_t) \propto a_t^{2.45 \pm 0.1}$ and $\sigma_{0,c} \approx (1.35\pm 0.40)\times 10^{-6} \mathrm{au}^2$. For Jupiter, we also find a dependence of $v_0(a_t)$ on $a_t$. It is given as $v_0(a_t) = ((15.44 \pm 1.10) \mathrm{km/s})a_t^{-0.36 \pm 0.02}$.

\begin{figure*}
    \begin{center}
    \includegraphics[width=1.0\textwidth]{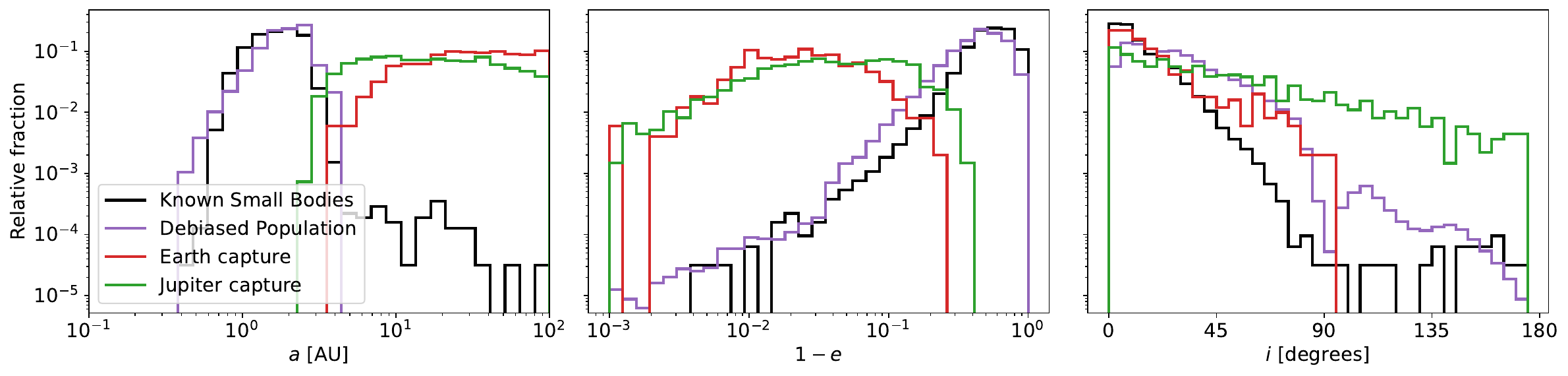}
    \caption{Distribution of known NEOs in the solar system and a debiased population of NEOs from \citet{Granvik2018Icar..312..181G} compared to distribution of captured bodies by Earth-Moon \& Jupiter using $v_{\infty}=8$km/s. We find that the distribution of captured bodies tends to be more eccentric and highly inclined compared to the distribution of NEOs (both detected and debiased) in the solar system. Long-term evolution of the captured population in necessary before making any concrete predictions. } 
    \label{fig:comparison_known}
     \end{center}
\end{figure*}

 Notably, objects captured by Jupiter have a different distribution of orbital parameters than those captured by Earth. In Figure \ref{fig:comparison_known}, we calculate the distribution of captured objects by Earth-Moon and Jupiter at $v_{\infty} = 8$ km/s and compare those to present day known NEOs along with a debiased population of 800k NEOs from \citet{Granvik2018Icar..312..181G}. The latter is necessary since since selection effects from observational surveys do not provide a proper representation of the true population.  We choose a larger $v_{\infty}$ for presentation since the hyperbolic excess velocity of any incoming ISOs in the present day solar system is expected to be large.  The median semi-major axes of captured objects with $a<100$ au by the Earth-Moon at $v_{\infty}=8$ km/s is 32 au, whereas that for objects captured by Jupiter is 16.5 au. Jupiter is also more efficient at capturing objects on retrograde orbits. In fact, the Earth-Moon system is unable to capture any objects on retrograde orbits at higher $v_{\infty}$. Jupiter is also able to capture certain objects on lower eccentricities than the Earth-Moon system. The minimum eccentricity of an object captured by Earth-Moon is 0.76, whereas that by Jupiter is 0.66.   

How do the orbital parameters of objects captured compare to the known population of NEOs today? 
Just for clarity, in Figure \ref{fig:comparison_known}, we have only considered a specific set of objects captured with $a < 100$ au at $v_{\infty}=8$ km/s. We find that observed NEOs have a median semi-major axis of 1.7 au  and are heavily skewed towards prograde and circular orbits. Most NEOs originate within the orbit of Jupiter. However, there is a smaller albeit significant fraction of NEOs originating from $a > 10$ au. Our simulations cannot generate any NEOs with $a < 2$ au. The object with the minimal semi-major axis captured by Earth is 4.5 au whereas that for Jupiter is 2.1 au. Objects cannot be captured on semi-major axes smaller than these values as that requires an encounter with the planet with periapsis smaller than the radius of the planet, resulting in a collision rather than capture. Although the debiased population is limited to objected with semi-major axis $<4.2$ au, we find similar results as that mentioned above. Interestingly, the debiased population contains a larger fraction of objects on highly eccentric or retrograde orbits compared to the observed population. Comparing the distributions, we find that, should an ISO were to be captured into an NEO orbit, it would more likely end up with $a > 10$ au. Incidentally, this is where the Centaurs exist. ISOs hiding amongst the Centaurs have been examined by \cite{Siraj2019ApJ...872L..10S} but no known Centaurs are considered to have an interstellar origin. However, our study suggests a closer examination may be merited. 
We caution the reader that this is not a full representation of present-day captured ISOs by Earth and Jupiter (should they exist). ISOs captured in the past will evolve their orbital parameters over time due to interactions with other planets in the Solar System. ISOs captured in the present day would also arise from a different velocity distribution which can affect the distribution of orbital parameters. 

\subsection{Long term evolution and survival of captured bodies}

\begin{figure}
    \includegraphics[width=0.5\textwidth]{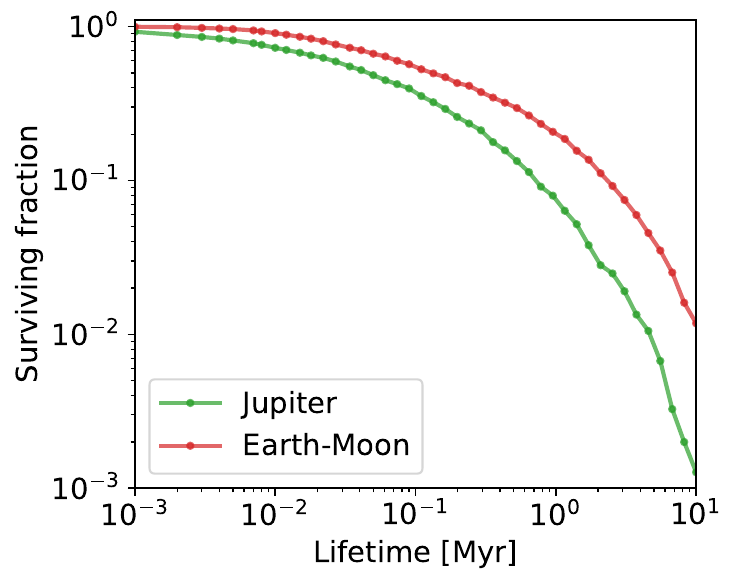}
    \caption{Survival fraction of captured interstellar NEOs as a function of the lifetime in bound NEO orbits. We notice that the survival fraction monotonically decreases and only 0.1-1\% of the captured objects survive in NEO orbits by 10 Myr. This is significantly shorter than the lifetime of known NEOs. Once leaving the NEO orbit, the object either becomes unbound or switches to an orbit with a larger perihelion and semi-major axis. Objects captured by Earth on average survive 2-3$\times$ longer than objects captured by Jupiter.} 
    \label{fig:survival_fraction}
\end{figure}

We perform long term evolution of a subset of captured objects in NEO orbits with $a<20$ au. The simulations are carried out until a duration of 10 Myr to understand how many objects survive over this duration. In total, we perform $\sim 20$k long-term simulations. The orbital parameters of the captured objects are stored for every 1000 years of evolution. 

We plot the survival fraction as a function of the lifetime of the captured interstellar NEOs in Figure \ref{fig:survival_fraction}. We note that the survival fraction only accounts for survival in NEO orbits. Thus, objects that are still bound but diffused into orbits with larger perihelion and semi-major axis are not taken into account. 

The survival fraction decreases monotonically and only 0.1-1\% of the captured objects survive in NEO orbits by 10 Myr. Although this is shorter than the bulk median lifetime of known NEOs ($\sim 10$ Myr), a closer analysis reveals that our results are consistent with NEOs originating from outer asteroid belt. \cite{Granvik2018Icar..312..181G} finds that the average lifetime of such objects is 0.3-0.4 Myr, which is consistent with our results. 
The captured objects, on account of being highly eccentric and prograde, tend to survive for shorter durations. Some objects collide with the planets or the sun during this time while other objects raise their perihelion upon close encounters with planets and move out of NEO orbits. 

Unlike \cite{Napier2021PSJ.....2..217N}, we stop our simulations at 10 Myr since we are mostly interested in understanding the characteristic lifetime of these captured objects. Due to the shorter duration, it is hard to ascertain what the functional form of the survival fraction is as a function of the lifetime. We do notice that at $t>1$ Myr, the survival fraction approaches a $\sim t^{-1.6}$ form for both objects captured by Earth and Jupiter. This is, in principle, similar to the findings of \cite{Napier2021PSJ.....2..217N} although we would need longer duration simulations to confirm this which are outside the scope of the current work.

We denote the time at which the surviving fraction is half the original fraction as $\tau_{0.5}$ and the time where the surviving fraction is one-tenth of the original fraction as $\tau_{0.1}$. Calculating $\tau_{0.5}$ from the data we find that $\tau_{0.5, jup} \approx 0.05$ Myr whereas $\tau_{0.5, earth} \approx 0.13$ Myr. We find $\tau_{0.1,jup} \approx 0.8$ Myr and $\tau_{0.1,earth} \approx 2.1$ Myr. We also calculate the average survival lifetime $\langle \tau \rangle$. In order to do so, we do not assume a functional form for the survival fraction but rather interpolate between the datapoints obtained from our simulations. To be conservative, we set the survival fraction to be $0$ for $t>10$ Myr while integrating to find the mean lifetime. We find that $\langle \tau \rangle_{earth} \approx 2.2$ Myr, whereas $\langle \tau \rangle_{jup} \approx 1.3$ Myr. Interestingly, it turns out objects captured by Earth survive longer. The survival fraction for captured ISOs by Earth is almost $10\times$ larger than that of objects captured by Jupiter at 10 Myr. We hypothesize that this might occur due to a larger fraction of Jupiter captured objects on chaotic, often retrograde orbits but a more thorough analysis is required which is beyond the scope of this study. 

The objects that are able to raise their perihelion beyond NEO orbits survive much longer. We noticed that the overall survival rate increases by as much as $10 \times$ when \textit{all} bound objects, regardless of whether they survive in NEO orbits, are taken into account. In that case, $\tau_{0.5}$ for all captured objects mirrors the results of \cite{Napier2021PSJ.....2..217N}. This is a noteworthy point since the objects that were simulated in \cite{Napier2021PSJ.....2..217N} had much larger semi-major axes. This suggests that close encounters in the solar system are often able to increase the semi-major axes of captured objects quite efficiently even for objects that start in really bound orbits of $a < 20$ au.

Although not presented here, we notice that the survival fraction is independent of initial $v_{\infty}$ of the ISOs. This is similar to the findings of \cite{Napier2021PSJ.....2..217N} where the authors found that the initial hyperbolic excess velocity played no role in determining the lifetime of captured objects.

\section{Discussion} \label{sec:discussion}
Using the capture cross section calculated in the previous sections, we now proceed to calculate the rate of capture of objects under various velocity distributions and object sizes. 

\subsection{Capture in present day solar system}

\begin{figure}
    \centering
    \includegraphics[width=0.5\textwidth]{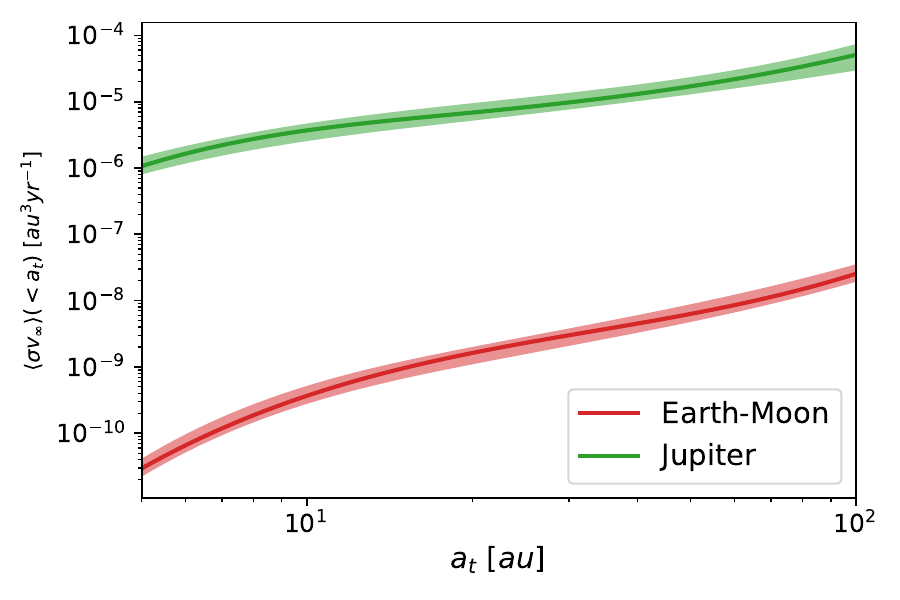}
    \caption{Velocity averaged cross-section of capture of ISOs ($\langle \sigma v_{\infty} \rangle$) as a function of the threshold semi-major axis $a_t$. The shaded region denotes the uncertainty arising from the uncertainty in the LSR velocities. The capture cross-section is presented for captures by both Earth-Moon and Jupiter. As noted in previous studies, we expect the capture rate to be dominated by Jupiter, even for NEOs. Jupiter is able to inject objects into NEO orbits at a rate that is $10^4\times$ that of Earth.} 
    \label{fig:comparison_cap_rate}
\end{figure}

To calculate the rate of capture, we first calculate the velocity averaged cross-section given as, 
\begin{equation}
    \langle \sigma v_{\infty} \rangle = \int_0^{\infty} f(v_{\infty}) \sigma(v_{\infty}) v_{\infty} 4 \pi v_{\infty}^2 dv_{\infty}
\end{equation}
where $f(v_{\infty})$ is the distribution function of the hyperbolic excess velocity, $\sigma(v_{\infty})$ is the functional form of the cross-section as a function of $v_{\infty}$. We assumed that the distribution of excess hyperbolic velocities is isotropic. We are interested in understanding rates of captures of objects due to close encounters with the Earth-Moon system and NEOs arising out of captures by Jupiter. To model the functional form for $\sigma(v_{\infty})$, we used an interpolation routine rather than assuming a fitted form. For values of $v_{\infty} > 15$ km/s for Earth-Moon and $v_{\infty} > 20$ km/s for Jupiter, we assumed $\sigma=0$ since we did not obtain any datapoints beyond these $v_{\infty}$ values. Thus, our calculated values of capture rates represent the lower bound.

In present day, we expect the hyperbolic excess velocity distribution of objects to resemble the field stellar velocity distribution. Accordingly, to estimate the rate of capture in present day solar system, we use the known values of the Sun's velocity relative to the Local Standard of Rest (LSR). The solar velocity relative to the LSR is given as $(v_U^{\odot}, v_V^{\odot}, v_W^{\odot}) = (10 \pm 1, 11 \pm 2, 7 \pm 0.5)$ km/s \citep[e.g.,][]{Schonrich2010MNRAS.403.1829S,Tian2015ApJ...809..145T,Bland2016ARA&A..54..529B}. The velocity dispersion of the local stars around the LSR is given as $(\sigma_U,\sigma_V,\sigma_W)=(33\pm4,38\pm4,23\pm2)$ km/s \citep[e.g.,][]{Anguiano2018MNRAS.474..854A}. Using the method described in \cite{Siraj2020ApJ...903L..20S}, we construct the probability distribution of incoming speeds of ISOs with respect to the Sun. For more information, we refer the interested reader to \cite{Siraj2020ApJ...903L..20S}. We note that we assumed the speeds are distributed isotropically, an assumption that might not hold in reality. 

We plot the velocity averaged cross section ($\langle \sigma v_{\infty} \rangle$) as a function of the maximum semi-major axis of captured objects ($a_t$) in Figure \ref{fig:comparison_cap_rate}. We find that $\langle \sigma v_{\infty} \rangle_{jup}/\langle \sigma v_{\infty} \rangle_{earth} \approx 10^4$. Thus the majority of the captured interstellar NEOs is due to Jupiter rather than Earth. It is a monotonically increasing function of $a_t$ indicating that captures
leading to orbits with larger semi-major axes are more likely than
those leading to orbits with smaller semi-major axes. 
Note that this is only for captures due to close encounters with the planets. Distant encounters do not play a role in capturing objects with $a<100$ au for both Earth and Jupiter and can be safely ignored.

Our results indicate that in the present day, capture is hard due to the relatively large velocity dispersion of field stars with respect to the Sun. A caveat of our results is the assumption that incoming ISO velocities reflect the field star velocities. However, gravitational kicks from planets lying in the habitable zone of M dwarf stars (the most abundant type of star) can eject objects at $v_e \sim 50$ km/s \citep[][]{Siraj2020ApJ...903L..20S}, where $v_e$ is the ejection velocity. If we assume that the ejection velocity is $50$ km/s, we find that $\langle \sigma v_{\infty} \rangle$ is almost 4$\times$ lower than that from before. This indicates the sensitivity of the results to the origin of ISOs and highlights the importance of modelling the origin of these objects properly.

Capture would have been possible in the solar birth cluster as well. Although we don't present a thorough analysis of capture in the birth cluster, assuming a functional form similar to that used in equation 34 in \cite{Napier2021PSJ.....2...53N} allows us to derive the approximate rates for capture in the solar birth cluster. The functional form for the ejection velocity is derived from \cite{Moorhead2005Icar..178..517M} and reflects the ejection due to a giant planet like Jupiter. Doing so, we find that $\langle \sigma v_{\infty} \rangle_{jup} (a_t = 10 \mathrm{au}) \approx 10^{-2} \mathrm{au}^3 \mathrm{yr}^{-1}$ which is almost $10^3 \times$ that in the present day environment. However, the configuration of the solar system in the birth cluster is a debated topic and the presence of nearby stars and the tidal field of the cluster would affect the capture and retention process. In light of these uncertainties, we defer a careful analysis of capture in the birth cluster to future studies.

\subsection{Present day population and prospects for detection}

To understand the present day population of interstellar NEOs, we calculate the rate of capture using the results from the previous section. The rate of capture ($\Gamma$) is defined as follows:
\begin{equation}
    \Gamma = n_{\mathrm{ISO}} \langle \sigma v_{\infty} \rangle
\end{equation}
where $n_{\mathrm{ISO}}$ is the number density of ISOs. Using 1I/`Oumuamua, the number density of Oumuamua sized objects has been determined to be $0.1-0.2 \mathrm{au}^{-3}$ \citep[e.g.,][]{Do2018ApJ...855L..10D,MoroMart2022ApJ...924...96M}. To estimate the size distribution of objects ($n_{\mathrm{ISO}}(d)$), we use the following normalization from \cite{Siraj2022ApJ...939...53S}:
\begin{equation} \label{eq:size_dist}
    n_{\mathrm{ISO}}(d) = n_{\mathrm{ISO},O} \left (\frac{d}{d_O} \right )^{1-q}
\end{equation} 
where $n_{\mathrm{ISO},O}$ is the number density of `Oumuamua sized objects, $d$ is the diameter of the object, $d_O$ is the diameter of `Oumuamua sized objects, and $q$ is a power-law index where $4<q<4.5$. To be conservative, we set $n_{\mathrm{ISO},O} = 0.1 \mathrm{au}^{-3}$ $d_O = 100$m. 
We find that the number density obtained from the normalization used above agree with those obtained from other works using double power law models \citep{MoroMart2022ApJ...924...96M}. In addition, \cite{Engelhardt2017AJ....153..133E} found an upper-limit of $0.00014 \mathrm{au}^{-3}$ for objects with $d > 1$ km. Using the above normalization, setting $d = 1$ km, we find that $n_{\mathrm{ISO}}(1 \mathrm{km}) \leq 10^{-4} \mathrm{au}^{-3}$, consistent with \cite{Engelhardt2017AJ....153..133E}. 

The number of ISOs ($N_{\mathrm{ISO}}$) in NEO orbits can then be estimated as
\begin{equation}
    N_{\mathrm{ISO}} = \Gamma \tau 
\end{equation}
where $ \tau $ is the lifetime of captured objects. To simplify our calculations and estimate the number of captured objects, we assume $\tau = \langle \tau \rangle$, the average survival lifetime, obtained in the previous section. Note that $\tau$ might vary depending on the semi-major axis of the captured objects but our assumption is valid in giving us an order-of-magnitude estimate on the current number of captured ISOs amongst NEOs.

\begin{figure*}
    \begin{center}
    \includegraphics[width=1.0\textwidth]{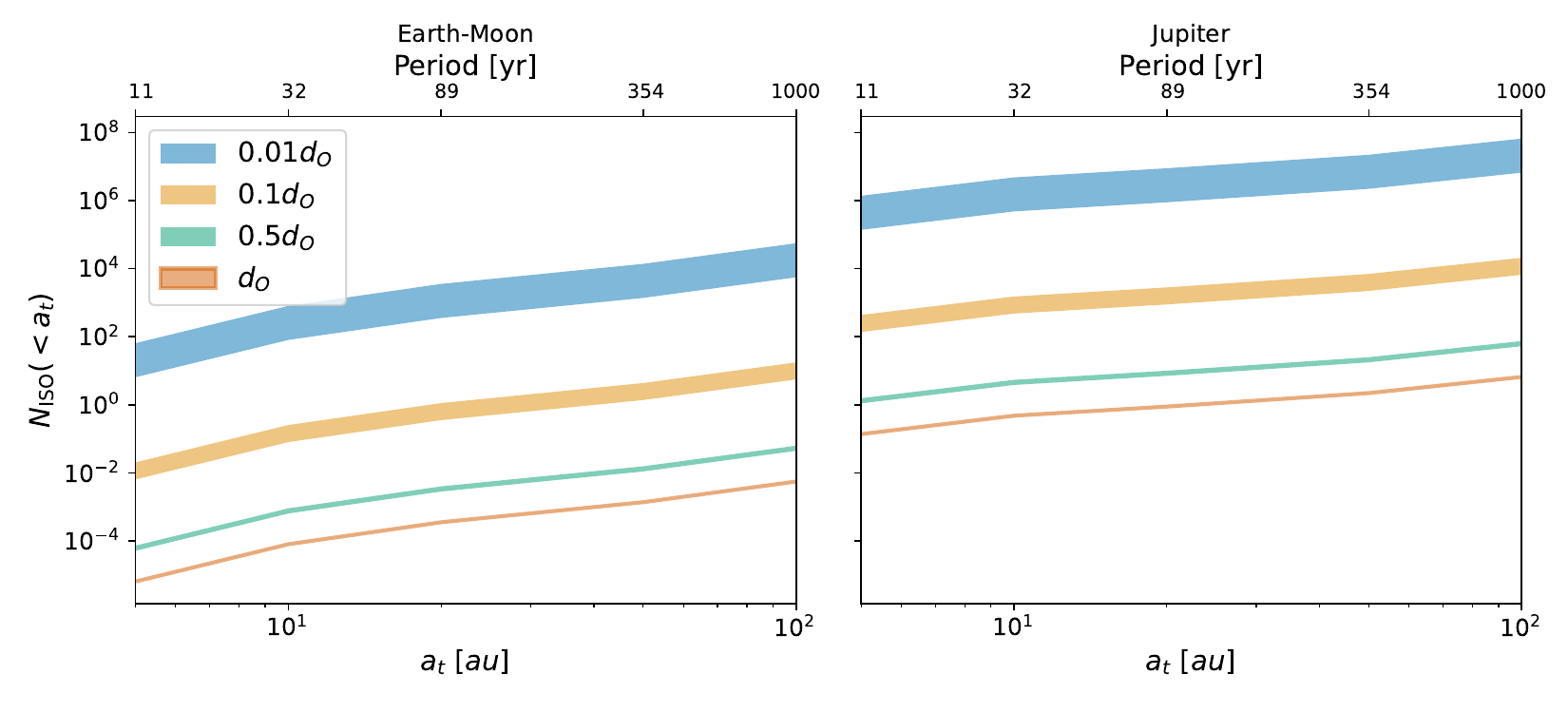}
    \caption{ Number of ISOs ($N_{\mathrm{ISO}}$) present in steady state as a function of the threshold semi-major axis $a_t$. The shaded regions represent the numbers calculated using different power law index parameters as given in equation \ref{eq:size_dist}. The lower limit is taken as $q=4$ and the upper limit is taken as $q=4.5$. \textit{Left:} captured ISOs as a result of close encounters with Earth-Moon. \textit{Right:} captured ISOs as a result of close encounters with Jupiter. Smaller sized objects ($\sim 1$ m) should outnumber larger sized objects. We expect $\mathcal{O}(10^6)$ of these objects to lie within the orbit of Saturn and in NEO orbits. $\sim 10$ m sized objects number are also present in NEO orbits today, according to our calculations. We expect to find a steady state population of $\mathcal{O}(10^3)$ such objects. We also find that about $5$ $\sim 50$ m sized captured objects with $a<10$ au should also be present amongst a population of NEOs today. This has major implications since these objects would be detectable by LSST over its lifetime. This calculation represents the estimate calculated using the mean survival lifetime. Under this estimate, we should not expect to find any 'Oumumua sized objects among NEOs today.} 
    \label{fig:comparison_cap_rate_today}
     \end{center}
\end{figure*}

We plot the estimated number of captured interstellar NEOs ($N_{\mathrm{ISO}}$) of different sizes by Earth and Jupiter as a function of the threshold semi-major axis $a_t$ using $\tau=\langle \tau \rangle$ in Figure \ref{fig:comparison_cap_rate_today}. Small interstellar objects of sizes $1/100$ that of `Oumumua ($\sim 1$ m) would form the dominant captured population owing to the large number density. We note that according to our calculations, a steady state population of $>10^5$ $\sim1$ m sized objects should be present with $a<5$ au among present day NEOs. The number increases to $5\times10^5$ for $a<10$ au. Our results also indicate that, although fewer, we would still expect a steady state population of $>500$ $0.1 d_O$ ($\sim 10$ m) sized objects among NEOs today with $a<10$ au. The steady state population drops fast with increasing diameter and we find that only 5-6 objects of size $0.5 d_O$ ($\sim 50$ m) should exist among NEOs today with $a<10$ au. Interestingly, we find that about 1 object this size should also exist with $a<5$ au. This has major implication for detectability as we discuss it in later. Unfortunately, we find that no Oumuamua sized objects exist among NEOs today.

The volume capture rate for captured ISOs was calculated to be $0.051 \mathrm{au^3} \mathrm{yr}^{-1}$ in \cite{Hands2020MNRAS.493L..59H}. Only 0.033\% of the total captured population was found to be lying within $a<6$ au. We find that using the same $n_{\mathrm{ISO}}$ and $\tau$ as above, it would imply a $N_{\mathrm{ISO}} \approx 2000$ for $\sim 10$ m sized objects with $a<6$ au. This is consistent with the $N_{\mathrm{ISO}}$ for similarly sized objects in our study indicating that our study is consistent with previous works.

\begin{figure*}
    \begin{center}
    \includegraphics[width=0.6\textwidth]{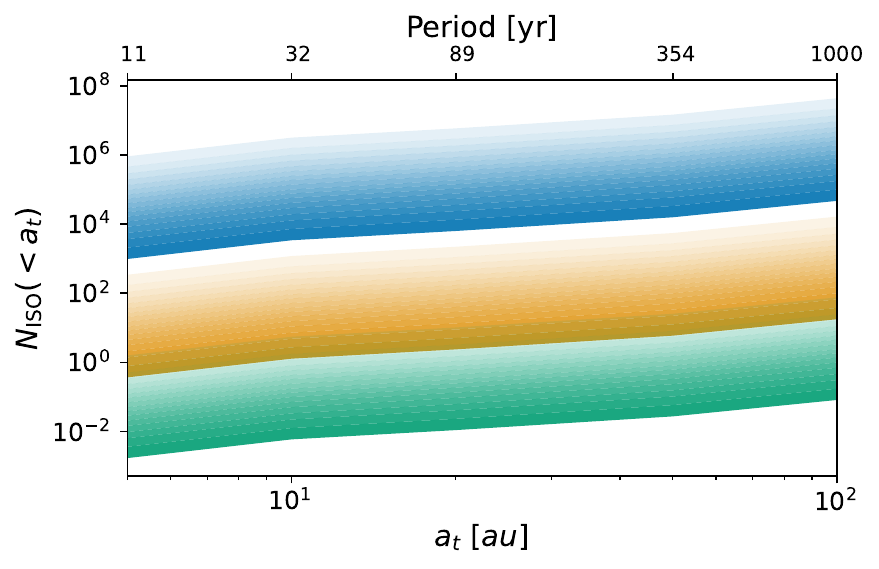}
    \caption{ Comparison of the number of captured ISOs in NEO assuming a survival lifetime based on the fraction of surviving population. The darkest shades denote the population estimates based on a timescale calculated using a 95\% survival fraction whereas the lightest shades denote a population calculated using a 5\% survival fraction. The colors denote the size of the object and are the same as that in Figure \ref{fig:comparison_cap_rate_today}. We find that a steady state population of smaller (1-10m) captured ISOs is highly likely to be present amongst NEOs today whereas population estimates for larger objects is more uncertain. An optimistic scenario is consistent with population calculated using the average lifetime and predicts a steady state population of 3-6 $\sim 50$ m in NEO with $a<10$ au. In a 10 year lifespan of LSST, we can assume to detect $\sim 1$ such object assuming that the brightness profile is similar to that of Oumuamua. }
    \label{fig:cap_rate_today_optimistic}
     \end{center}
\end{figure*}

Since the number is highly sensitive to the lifetime of these captured objects, we perform a similar analysis as above where we use lifetimes corresponding to various survival fractions and present them in Figure \ref{fig:cap_rate_today_optimistic}. The darkest shade indicates the population calculated using a 95\% survival fraction whereas the lightest shade indicates a population calculated using a 5\% survival fraction. We notice that the steady-state population numbers calculated using this method are consistent with the values calculated above using the average lifetime. Importantly, it highlights the dependence on the present day population and hence detectability on the lifetimes of such objects in NEO orbit. We find that a steady state population of $\sim 1-10$ m sized captured ISOs is highly likely to present today. Using a lifetime corresponding to the half-life of the population, we find that $>10^4$ 1m sized objects and $>10^2$ 10 m sized objects would be present amongst NEOs today with $a<10$ au. The population statistics for larger objects is more uncertain with 0-10 $\sim 50$ m sized objects forming a steady state population today. Future studies should focus on careful modelling on the lifetime of such objects and include non-gravitational forces since they can significantly affect how long these objects survive.

We consider the detectability of such objects using the Vera Rubin Observatory (LSST). LSST has a limiting magnitude of $\sim 24.5$ in a 30 second exposure. Assuming an albedo similar to that of `Oumuamua, we find that the absolute magnitude of $50-60$ m sized objects in NEO should be around the limiting magnitude of LSST. Assuming these objects have $a\sim 10$ au would imply a period of rotation of $\sim 30$ years. This would imply that over a 10 year run of LSST, we should be able to detect $\sim N_{\mathrm{ISO}} \frac{t_{\mathrm{LSST}}}{t_{\mathrm{ISO}}} = \frac{1}{3} N_{\mathrm{ISO}}$. Using the estimate for $N_{\rm ISO}$ derived using the average lifetime implies that LSST should be able to detect $\mathcal{O}(1)$ such captured interstellar NEOs during its lifetime. If these captured objects had larger albedo, LSST would be able to detect even smaller objects going down to $20-30$ m sized objects enabling 10-100 detections over its lifetime. Similar analysis can also be performed with the Hyper Suprime Cam (HSC) which gives us prospects of detectability close to those obtained using LSST.

It is evident from our study that smaller captured ISOs outnumber larger ISOs. However, detecting these $1-10$ m sized objects is a hard task. \cite{Shao2014ApJ...782....1S} propose synthetic tracking to detect very small objects down to sizes of 7 m among NEOs. Such a method would allow detections of smaller interstellar NEOs and enable $\mathcal{O}(10)$ detections per year. Machine-learning (ML) based methods \citep[e.g.,][]{Hefele2020A&A...634A..45H} may also improve the ability to detect smaller objects. We plan on investigating ML based techniques for detection in the future. Prospects for detectability using future space based observatories such as NEO Surveyor should also be investigated in the future. The detections should focus on objects with orbits that are largely prograde and highly eccentric although work needs to be done to properly ascertain the distribution of orbital parameters after the object has undergone encounters with planets in the solar system after getting captured. This is beyond the current scope of the study and will be investigated in the future.

\section{Conclusions} \label{sec:conclusions} 
Interstellar objects present a unique mechanism to investigate the formation and evolution of planetary systems including our own. Although rare, ISOs can be captured into bound orbits by different planets in the solar system. Whether any captured ISOs exist in th present-day solar system is a major area of interest. A population of captured ISOs can provide valuable information regarding the origins of such objects.

In this study we examined the capture of ISOs into near Earth orbits using a large suite of $N$-body scattering simulations using a new hybrid integration scheme. We computed the capture cross-section of Earth-Moon and Jupiter and found that Jupiter should dominate the capture of interstellar objects into near Earth orbits by a factor of $10^4$ compared to that of Earth-Moon. These captured objects have larger semi-major axes, lower inclinations, and higher eccentricities than the current known population of Near Earth Objects. We find that most captured interstellar NEOs in the solar system arise from beyond the orbit of Saturn.

Long term simulations of such objects reveals that they're  chaotic and have an average survival lifetime of $\sim 1$ Myr. We find that the average survival lifetime of such objects in NEO orbits is $\approx 1.3$ Myr for objects captured by Jupiter and $\approx 2.2$ Myr for objects captured by Earth. Although this is lower than the bulk lifetime of known NEOs, closer analysis reveals that the survival lifetime is consistent with NEOs arising out of the outer asteroid belt. 

The computed cross section is then used to determine the capture rate in the present day solar system assuming the ISOs originate with a velocity distribution similar to that found for the field stars in the solar neighborhood. The capture rate is computed for objects of different sizes. Using the computed life time of survival, an approximate number of ISOs among NEOs in the steady state is determined.

We find that smaller ($\sim 1$ m) sized objects should dominate the captured population and can make up a population of $> 5\times 10^5$ objects with $a<10$ in NEO orbit. $\sim 10$ m sized objects should be present with a population of 100-1000. Our simulations show about 5 $\sim 50$ m sized objects should be present with $a<10$ au. Unfortunately, according to our simulations, no Oumuamua sized objects are able to be captured and retained in NEO orbit. However, such objects may exist within the solar system but not in NEO orbits. 

The captured population can have major implications for detectability. We predict that LSST should be able to detect 1-2 $\sim 50$ m sized captured ISOs among NEOs over its lifetime. Although LSST would only be able to resolve $\sim 50$ m sized objects with $a<10$ au, other mechanisms such as synthetic tracking should be able to detect objects as small as 7 m among NEOs. Surveys should focus on objects with orbits that are highly prograde, eccentric, and have semi-major axes of $a\sim10$ au. However, to better ascertain the distribution of orbital parameters, long term simulations should be used.

To our knowledge, this is the first study that explores the possibility of presence of interstellar objects in our neighborhood among NEOs. Interstellar NEOs would be preferable to other captured ISOs due to the proximity to Earth which would aid in detectability and possible visits by space probes. With a substantial number of potentially detectable objects hiding among NEOs, we demonstrate the need and merit for more research in the areas of dynamics and detection of these small-sized objects in the future.

\section*{Acknowledgements}

 We thank the anonymous referee whose comments substantially improved the paper. We acknowledge the usage of the Vera cluster, which is supported by the McWilliams Center for Cosmology and Pittsburgh Supercomputing Center. DM acknowledges support from the McWilliams Center-Pittsburgh Supercomputing Center seed grant and NASA grant 80NSSC22K0722.
HT acknowledges support from NASA grants 80NSSC22K0722 and 80NSSC22K0821, and NSF grant PHY2020295. 

\section*{Data Availability}

The data from the $N$-body simulations and the code used for this work are available upon reasonable requests.



\bibliographystyle{mnras}




\appendix

\section{Comparison to IAS15}

\begin{figure}
    \centering
    \includegraphics[width=0.45\textwidth]{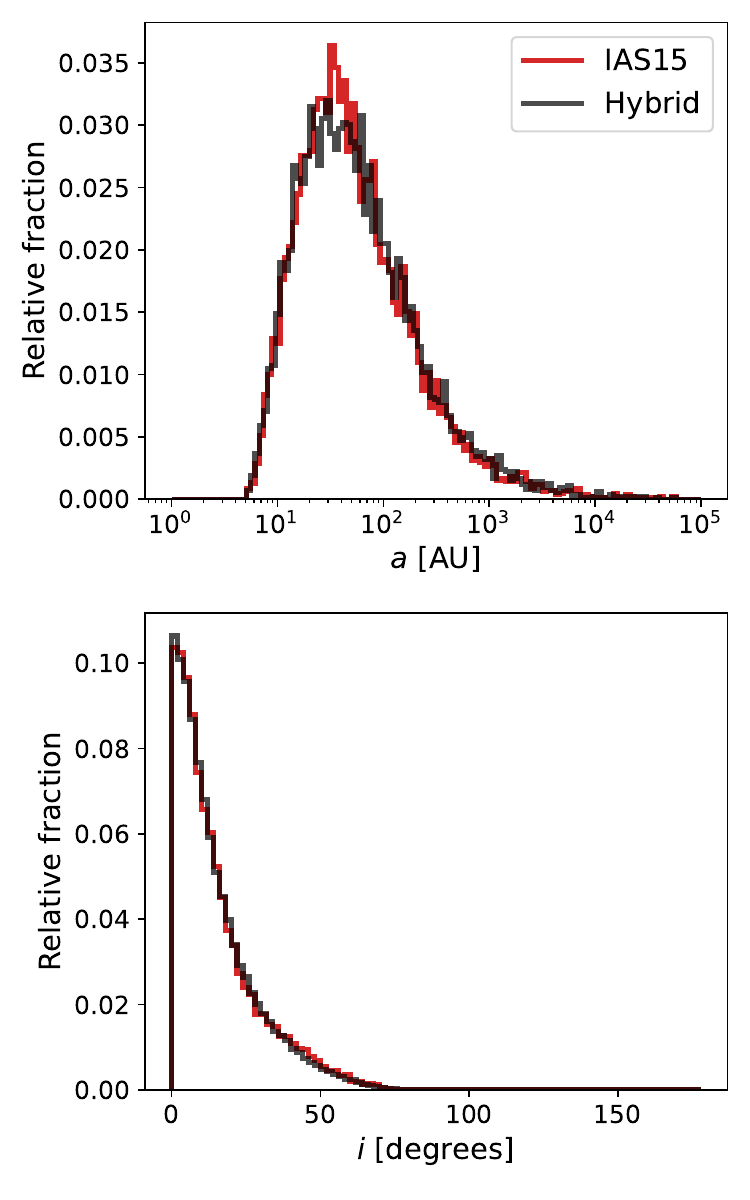}
    \caption{Distribution of the orbital parameters of captured objects for $v_{\infty}=8$ km/s using the IAS15 integrator and our hybrid method. We notice that the distributions of semi-major axis ($a$) and inclination ($i$) of the captured objects are very similar between the two schemes. This validates the robustness of our hybrid method.}
    \label{fig:ias15_comparison}
\end{figure}

In order to verify the validity of our hybrid integration scheme, we run a set of $\sim 10^7$ scattering experiments using the IAS15 integrator \citep{Rein2015MNRAS.446.1424R} and compare the orbital parameters of the captured objects to those obtained using our hybrid scheme. For the sake of clarity, we only present the results from $v_{\infty}=8$ km/s. We plot the distribution of semi-major axis ($a$) and inclination ($i$) obtained from using IAS15 and our hybrid method in Figure \ref{fig:ias15_comparison}. We notice that the distributions of captured orbital parameters is statistically similar using both methods. The median semi-major axis of the captured objects using IAS15 is $44.6$ au while that using our hybrid method is $46.2$ au. The 10 and 90 percentile values of the captured semi-major axis using IAS15 are $13.1$ au and $299.8$ au respectively. Using our hybrid method, we obtain $13.2$ au and $322.8$ au for the 10 and 90 percentile values of the captured semi-major axis respectively. The median inclination using IAS15 was found to be $11 \degree$ whereas that using our hybrid method was found to be $10.5 \degree$. This indicates that the results obtained using our method are statistically similar to those using IAS15, confirming the validity of our simulations. Although not presented here, we performed a similar analysis with
smaller values of $v_{\infty}$ and found similar results thereby confirming that our method is able to simulate both weak and close encounters properly.


\bsp	
\label{lastpage}
\end{document}